\documentclass[a4paper,fleqn,usenatbib]{mnras}
\usepackage[T1]{fontenc}
\usepackage{ae,aecompl}

\usepackage{graphicx}	
\usepackage{amssymb}	

\usepackage{color}

\newcommand{\msun}{{\rm M}_{\odot}}

\title[Correlated QPOs in SWIFT~J1753.5--0127]
      {Discovery of correlated optical/X-ray quasi-periodic oscillations in black hole binary SWIFT~J1753.5--0127}
 \author[Veledina et al.]
    {Alexandra~Veledina,$^{1,2}$\thanks{E-mail: alexandra.veledina@gmail.com} 
     Mikhail~G.~Revnivtsev,$^{3}$ 
     Martin~Durant,$^{4}$ Poshak~Gandhi,$^{5}$
      \newauthor and Juri~Poutanen$^{1}$
  \\
$^1$Tuorla Observatory, Department of Physics and Astronomy, University of Turku, V\"ais\"al\"antie 20, FI-21500 Piikki\"o, Finland\\
$^2$Nordita, KTH Royal Institute of Technology and Stockholm University, Roslagstullsbacken 23, SE-10691 Stockholm, Sweden\\
$^3$Space Research Institute, Russian Academy of Sciences, Profsoyuznaya 84/32, 117997 Moscow, Russia\\
$^4$Department of Medical Biophysics, Sunnybrook Hospital M6 623, 2075 Bayview Avenue, Toronto M4N 3M5, Canada \\
$^5$School of Physics and Astronomy, University of Southampton, Highfield, Southampton SO17 1BJ, UK
}

\date{Accepted 2015 September 20.  Received 2015 September 20; in original form 2015 August 8}

\pubyear{2015}

\begin{document}
\label{firstpage}
\pagerange{\pageref{firstpage}--\pageref{lastpage}}
\maketitle

\begin{abstract}
We report the discovery of the correlated optical/X-ray low-frequency quasi-periodic oscillations 
(QPOs) in black hole binary SWIFT~J1753.5--0127.
The phase lag between two light-curves at the QPO frequency is close to zero.
This result puts strong constraints on the nature of the optical emission in this object and on the 
origin of the QPOs in general.
We demonstrate that the QPO signal and the broadband variability can be explained in 
terms of the hot accretion flow radiating in both optical and X-ray bands.
In this model, the QPO appears due to the Lense-Thirring precession of entire flow, while the 
broadband variability in the optical is produced by two components: the hot flow and the irradiated disc.
Using the phase-lag spectra, we put a lower limit on the orbital inclination $i\gtrsim50$\degr, 
which can be used to constrain the mass of the compact object. 
\end{abstract}


\section{Introduction}

Accreting black holes (BHs) in our Galaxy have been at a close look by the X-ray telescopes for the last half a century.
Most of them are transient sources, typically being active for a few months and spending years to decades in quiescence 
\citep{RM06}.
X-ray studies revealed that the objects go through several spectral states during the outburst, classified according to the 
energy range where most of the energy is liberated \citep{ZG04}.
BH X-ray binaries demonstrate a tight connection between the spectral states and the variability properties \citep[see e.g.][]{GCR99,ABL05}.
The objects are highly variable on the sub-second time-scales in the hard state and this variability is largely suppressed
when it enters the soft state \citep[see review in][]{DGK07}.

The accretion geometry close to the compact objects and physical mechanisms shaping their broadband spectra are among 
the major questions still being debated.
Analysis of X-ray spectral and timing properties are the powerful tools to probe the physical processes 
operating in the vicinity of compact object.
The shape and the magnitude of the power spectra are connected to the spectral states and, ultimately, to the accretion geometry.
In the hard and hard-intermediate states, the variability spectra of BHs are characterised by a high amplitude broadband noise
(rms amplitude of about 30~per~cent), in addition to which narrow features
known as quasi-periodic oscillations (QPOs) are often detected \citep{BS14}.
Hereafter, the QPOs refer to the low-frequency QPOs observed in the frequency range $\sim$0.1--10~Hz, 
with the central frequency correlated to the low-frequency break of the broadband noise and the X-ray flux \citep{WvdK99,CBS05}.
The whole set of the QPO models was developed, yet the connection of these timing features to spectral formation has not been 
addressed in most of them.
The most established in this respect is the model where QPOs appear due to the Lense-Thirring precession of a hot accretion 
flow \citep{SV98,FB07,IDF09}.

Lately, intriguing properties of the fast optical variability and its correlation with the X-rays in Galactic BHs with low-mass 
companions also captured attention of the community.
This variability is likely produced in the processes of accretion/ejection and thus can also serve as a probe of the accretion physics.
The most straightforward method to find the connection between optical and X-rays is to calculate the cross-correlation 
function (CCF).
Three transients, GX~339--4, XTE~J1118+480 and SWIFT~J1753.5--0127 were found to demonstrate 
an anti-correlation between light-curves in these energy bands \citep[the so-called precognition dip 
in CCF]{Motch83,Kanbach01,HBM09,DGS08,GMD08}.
These observations immediately ruled out the reprocessing nature of the optical photons, favouring their non-thermal origin, 
in the hot accretion flow or, perhaps, the jet \citep[see e.g. recent reviews by][]{PV14,UC14}.
Interestingly, many observations also revealed the presence of the QPOs in the optical, ultraviolet and infrared power spectra at 
frequencies similar to those in the X-rays \citep{HHC03}.
So far, the relation of the optical broadband variability as well as the QPOs to the noise and low-frequency QPOs in the X-rays 
was not studied.

In this paper we study the variability patterns of the BH binary SWIFT~J1753.5--0127.
We use the data obtained in 2007, when the source was in the hard state.
The simultaneous observations report the presence of the QPO in the optical power spectra, but no statistically significant QPO 
in the X-rays was detected \citep{DGS09}.
Using the CCF and the phase-lag spectra, we show that the optical QPO has a coherent twin in the X-rays.
We calculate phase lags between the light-curves in the optical and X-rays
and demonstrate how this information can be used to determine the system inclination.
 
The remainder of the paper is laid out as follows. In Section~\ref{sect:data_analys} 
we present our data analysis and cross-correlation results; 
in Section~\ref{sect:model} we introduce a physical model which can describe the simultaneous broadband variability seen in both X-rays and the optical and their coupled QPOs. 
 We present our conclusions in Section~\ref{sect:conclus}.

\begin{figure}
\centering 
\includegraphics[width=6cm]{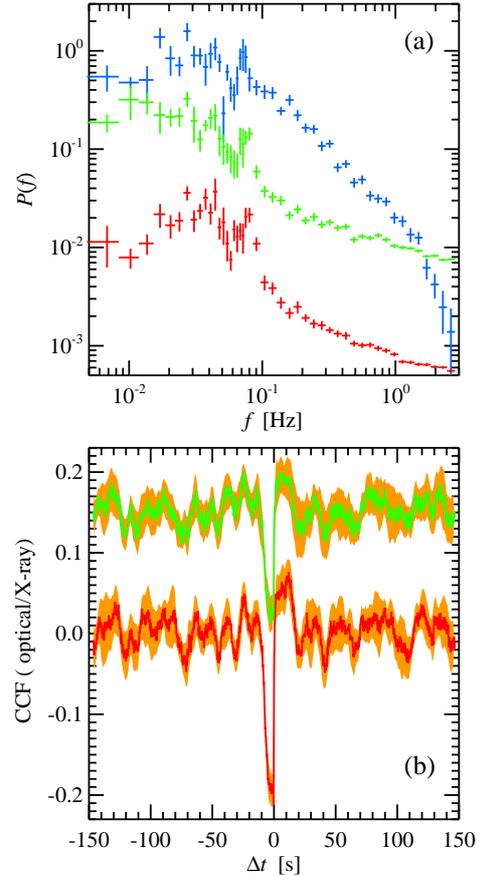} 
\caption{
(a) X-ray (upper, blue), $g'$ (middle, green) and $r'$ (lower, red) PSDs.
Poisson noise has been subtracted in the X-ray PSD.
There is a prominent optical QPO and a hint of the X-ray QPO is seen at the same frequency.
Analysis of \citet{DGS09}, however, suggests that the X-ray QPO is not statistically significant.
(b) $r'$/X-ray (lower) and $g'$/X-ray (upper, shifted up by 0.15 for clarity) CCFs. 
Orange shaded area reflects $1\sigma$ errors.
An oscillating structure is apparent at $|\Delta t|>10$~s.
Positive time-lags correspond to optical delay. 
}\label{fig:psd_ccf} 
\end{figure}

\section{Data analysis}\label{sect:data_analys}

\subsection{Data processing}

We reanalyse the data obtained on 2007 June 13 with VLT/ULTRACAM \citep{Dhillon07} simultaneously with
\textit{RXTE}/PCA.
The data were described and published in \citet{DGS09}.  
Optical light-curves were obtained in two filters $r'$ and $g'$, with 0.143~s time resolution.
Although the $u'$ light-curve was also obtained, we do not use it because poor weather conditions during the observations
resulted in a high noise level.
X-ray data were analysed with the help of package {\sc heasoft}.
We extracted the X-ray light curve applying solar system barycenter correction ({\tt faxbary}) 
and selected only X-rays in the energy band 3--15~keV.
The X-ray light-curve has the same 0.143~s time resolution.
 
We start our analysis of the optical light-curves with subtraction of the linear function fitted to the raw data.
This procedure reduces the noise, which could affect the calculated characteristics, however, it
does not affect principal results, which would have been the same even using the raw data.
Because of a variable sky transparency during the observations, we used the flux ratio of the 
target to the field star to constitute the optical light-curve.
This procedure aims to reduce the low-frequency multiplicative atmospheric noise, which is the same for 
the target and the comparison star.
It, however, does not eliminate the contamination of the atmospheric noise at higher frequencies corresponding to 
separation between the target and the comparison star.
The atmospheric noise is clearly seen in the optical power spectral density (PSD).
At some frequencies it may dominate over the source variability making it difficult 
to separate the intrinsic source variability from the noise contamination.
We use arbitrary normalisation for the PSDs for the same reason and further consider only the PSD shape.

\begin{figure*} 
\centering 
\includegraphics[width=12cm]{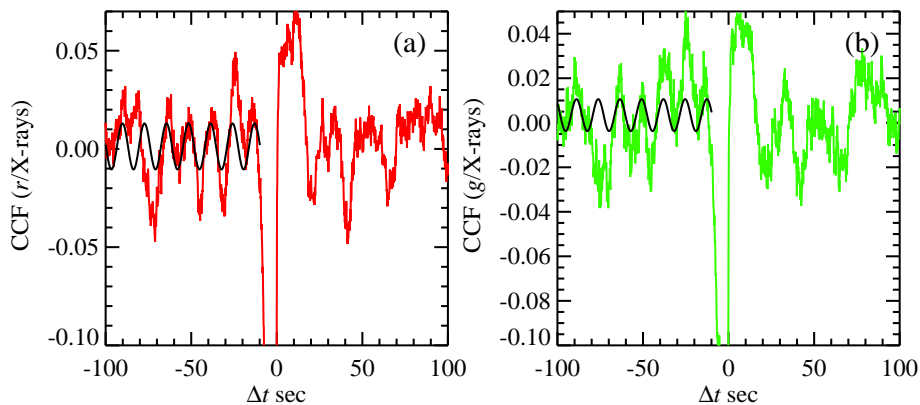} 
\caption{The observed CCF (a)  $r'$/X-ray and (b) $g'$/X-ray and 
the corresponding best-fitting cosine function.
}\label{fig:ccf_fit_cos} 
\end{figure*}

\subsection{Results}

We split the light-curve into 10 segments of about 290~s and calculate PSD in every segment.
The errors are estimated using bootstrap method.
We run $10^6$ simulations to compute average PSDs from randomly-chosen 10 segments 
(each segment can be repeated), and calculate the dispersion of these average values.
In Fig.~\ref{fig:psd_ccf}(a) we show optical and X-ray PSDs.
We focus on the variability in the frequency range $\sim$0.01--0.4~Hz.
We note the excess in the optical power at frequency $\sim$0.08~Hz 
\citep[as reported in][]{DGS09},
and some excess in the X-ray PSD at a similar frequency. 
Although \citet{DGS09} state that the QPO is not statistically required in the X-rays.

We calculate the CCFs in each segment by calculating the inverse Fourier transform of the cross-spectral density (CSD).
The errors are again calculated using bootstrap method.
We plot the optical/X-ray CCFs in Fig.~\ref{fig:psd_ccf}(b).
Positive time-lags correspond to optical delay.
Similar to \citet{DGS08}, we obtain an anti-correlation of the two light-curves at small ($-10\lesssim\Delta t\lesssim0$~s) 
negative lags, accompanied by positive correlation at similarly small positive lags.
We note that this dip-plus-peak structure is seen in all the time segments, as well as in the dynamical CCF 
\citep[figure 4 of][]{DSG11}.
In addition, we notice the oscillating structure at $|\Delta t|\gtrsim10$~s suggesting a distinct frequency is present 
in the optical and X-rays throughout the light-curve.
Both signals are correlated at this frequency and the correlation holds for a number of periods.

We fitted the CCF with the cosine function in the interval $-100<\Delta t<-10$~s, with frequency, phase, constant level 
and amplitude being free parameters.
The results are plotted in Fig.~\ref{fig:ccf_fit_cos}.
The best-fitting phase for both optical filters was found to be zero within the chosen accuracy.
The best-fitting oscillation frequency is $f=0.078$~Hz for $r'$/X-ray CCF, with the reduced $\chi^2$/d.o.f.=362.6/627, 
very close to the QPO frequency found by \citet{DGS09} in the optical PSD.
The $g'$/X-ray CCF shows oscillation at similar frequency 
of $f=0.079$~Hz with $\chi^2$/d.o.f.=404.8/627.
The reduced $\chi^2$ being less than unity in both cases means that the errors are somewhat overestimated.
The fact that the obtained frequencies are close to the optical QPO frequencies gives us grounds to suspect that the QPO 
is also present in the X-ray light-curve, but is veiled there under the broadband noise.
The significance of this result is, however, difficult to estimate.

\begin{figure}
\centering 
\includegraphics[width=6cm]{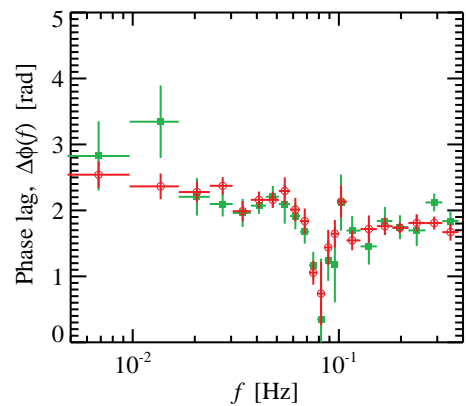}  
\caption{
Phase lags between the optical and X-ray light curves as a function of Fourier frequency.  
The error bars were obtained using equation (9.52) of \citet{BP86}.
Green squares and red circles with error bars show phase lags from the $g'$ and $r'$ filters compared to the X-ray light-curve, respectively. 
The phase-lags are almost independent of frequency with $\Delta\phi\approx2$~rad at $f\lesssim0.05$~Hz and 
decrease down to $\Delta\phi\sim 0.5$~rad at the QPO frequency.
}\label{fig:phase_lags_data} 
\end{figure}

If the QPO is present in both optical and X-rays, this should be reflected in their 
combined characteristics: the CCF and the phase lags.
We checked whether the phase-lag spectra demonstrate any statistically significant feature at the QPO frequency.
This time, the total light-curve was split into 20 segments, about 147~s each, as we are more interested 
in having more segments to improve statistical properties rather than in good frequency resolution.
Smooth otherwise, phase lags in both filters show an abrupt drop at the QPO frequency by about 1.5~rad 
(down to $\Delta\phi\approx0.5$~rad), with the width of the dip of only $\sim$0.02~Hz.
To assess the significance of the dip, we calculate the errors in the phase-lag spectra using a bootstrap method 
and from equation (9.52) of \citet{BP86}.
The errors calculated by both methods are in good agreement in general, however sometimes the former method 
gives larger errors due to the $2\pi$ phase lag uncertainty.
In Fig.~\ref{fig:phase_lags_data} we plot the average phase lags and their standard deviations calculated using 
the latter method. 
We calculate the average phase lag over the frequency range 0.008--0.2~Hz, excluding the frequencies 0.06--0.1~Hz,
where the QPO is likely contributing, as
\begin{equation}
 \Delta \phi_{\rm ave} = \frac{\sum\limits_{\rm i}^{} \Delta \phi_{i}/\sigma_{\rm i}}{\sum\limits_{\rm i}^{} 1/\sigma_{\rm i}},
\end{equation}
where $\Delta \phi_{i}$ and $\sigma_{\rm i}$ are the phase-lag and its standard deviation at $i$-th frequency.
We obtain $\Delta \phi_{\rm ave}=1.9$~rad for $r'$/X-ray  and $\Delta \phi_{\rm ave}=2.2$~rad for $g'$/X-ray  phase-lags.
The significance of the dip is $4.7\sigma$ for $r'$ band and it is $3.2\sigma$ for $g'$ band.
We conclude that the dip at the QPO frequency is statistically significant in both bands.
Further analysis is applied only to $r'$-filter as an illustration.

Our findings can be summarised as follows.
The optical data of SWIFT~J1753.5--0127 reveal a number of puzzles: there is a prominent QPO in the optical 
and only a hint of it in the X-ray PSD, a dip-plus-peak behaviour of CCF at small lags and its oscillating structure with 
the period close to the QPO period at larger lags, and the phase lags $\Delta\phi\cong2$~rad, almost independent of frequency 
with an abrupt drop down to nearly zero at the QPO frequency.

\section{Modelling}\label{sect:model}

Below we develop a quantitative model to explain the features revealed by the simultaneous data. 
Any successful model should simultaneously describe the broadband noise both in the X-ray and the optical band, 
the non-trivial relation between them reflected in a complex shape of the CCF and the phase-lags, 
as well as the presence of the QPOs in both bands with the nearly zero phase lag. 
The model should also be consistent with the known spectral characteristics of accreting BHs.

There is a general agreement that the X-rays in accreting BHs depending on the spectral state are produced by the 
inner hot accretion flow surrounded by a cold  disc and/or hot non-thermal corona above the disc 
\citep[see][ for reviews]{P98,ZG04,DGK07,Gilfanov10,PV14}. 
There could be at least three sources of optical radiation: the outer irradiated accretion disc, the inner hot flow, 
and the jet \citep{PV14}. 
It is likely that in  SWIFT~J1753.5--0127 contribution of the jet is negligible given 
a very low level of radio emission \citep{Soleri10,2015ApJ...808...85T}. 
Earlier we have developed a synchrotron self-Compton--disc reprocessing model that successfully described  
the X-ray/optical broadband timing properties. In this model, the X-rays were produced by the inner hot flow,
while the optical emission had contribution from two components: the synchrotron radiation 
from non-thermal electrons in the hot flow and the reprocessed emission from the outer cold disc. 
The synchrotron component was assumed to be anti-correlated with the X-rays with no delay; 
such a behaviour is expected in a hot flow, when the increasing mass accretion rate leads to 
increasing synchrotron self-absorption in the sources. The reprocessed disc emission is expected 
to be delayed, somewhat smeared but positively correlated with the X-rays.  
 
The model  described above  does not consider the QPOs at all.  
Recently a physically realistic model of Lense-Thirring precession of the whole hot flow 
that describes well the X-ray QPOs was developed \citep{FB07,IDF09}. 
\citet{VPI13} have pointed out that if the hot flow emits optical photons by synchrotron radiation, then 
the hot-flow precession will naturally produce not only  X-ray but also optical QPOs, 
either coming in phase or anti-phase,  depending on the observer position. 

The model we put forward is thus based on the combination of the 
synchrotron self-Compton--disc reprocessing model with the Lense-Thirring precession model of the hot flow. 
In this model,  there are two distinct classes of variability: the broadband noise coming from the accretion rate fluctuations and the 
(quasi-)~periodic signal due to a rotating emission pattern of the hot flow.
We consider the hot flow as a precessing torus, radiating both the X-rays and the optical emission.
We see more emission at both wavelengths when the observed surface area is maximal,
resulting in the QPOs coming in phase.
Simultaneously, the torus is getting brighter and fainter in an aperiodic manner depending on the accretion rate and 
independently of the precession phase.
The brightening in the X-rays is accompanied by fading of the optical synchrotron emission 
as proposed by \citet{VPI13}, resulting in the anti-correlation of the broadband noise. 
The mathematical description of this complex variability patterns is given in the following sections.

\subsection{Broadband variability}

For the description of the correlated optical/X-ray broadband variability we 
adopt the  two-component model described in \citet{VPV11}. 
The optical emission is assumed to be coming from two terms: from the synchrotron emission of the hot 
accretion flow which is anti-correlated with the X-rays and from the reprocessed emission which is 
correlated, but delayed and smeared with respect to the X-rays.
We also now take into account the fact that the synchrotron emission is coming from the larger radii
compared to the X-rays \citep{VPV13}, thus it lacks the high-frequency signal generated at smaller radii.

We assume that the X-ray variability is caused by propagating mass accretion rate fluctuations $\dot{m}(t)$ \citep{Lyub97}.
We assume the efficient accretion and 
take the X-ray light-curve $x(t)\propto\dot{m}(t)$, thus the broadband X-ray variability has the same PSD as the mass 
accretion rate PSD.
We use the zero-centred Lorentzian formalism \citep{NWD99} to describe its shape.
We denote the Fourier image of mass accretion rate by  $\dot{M}(f)$, where $f$ is the Fourier frequency, and use small letters with 
argument $t$ to denote variables in time domain and capital letters with argument $f$ for those in the frequency domain hereafter.
Mass accretion rate time-series are simulated from the observed X-ray PSD using the \citet{TK95} algorithm, with zero mean.

The optical synchrotron emission $s(t)$ is anti-correlated with the X-rays, but in addition it does not have 
the high-frequency signal because it is likely produced in the outer part of the hot flow.
This is simulated by multiplying $\dot{M}(f)$ by a filter function 
\begin{equation}\label{eq:filter_func}
 H(f) = \frac{1}{(f/f_{\rm s})^4+1},
\end{equation}
which describes the damping of high-frequency fluctuations of the hot flow above the characteristic frequency $f_{\rm s}$. 
We thus define the filtered mass accretion rate $\dot{m}_{\rm filt}(t)$ as the inverse Fourier transform 
of the product  $\dot{M}(f)H(f)$.  
In addition, the synchrotron emission can be coming earlier than the X-rays.
This can be controlled by introducing time-lag of the X-ray radiation compared to synchrotron, we find,  however, that 
no significant delays are required by the data, thus we further assume them to be zero.

The disc reprocessed emission is simulated 
as a convolution of the accretion rate and the response function $d(t)=\dot{m}(t) * r(t)$, where $r(t)$ is given by 
\begin{equation}\label{eq:response}
  r(t) = \left\{
 \begin{array}{cc}
  \exp\left[ -(t-t_1)/t_2 \right]/t_2 , & t \geqslant t_1, \\
  0,                                             & t < t_1. 
 \end{array} \right.
\end{equation}
It has two parameters: delay $t_1$ and the decay time $t_2$.
The corresponding Fourier image of the disc emission is given by
\begin{equation}\label{eq:disc}
 D(f) = R(f)\dot{M}(f).
\end{equation}

\subsection{Coupling with QPO}

The QPO can be obtained from the periodic signal by varying either amplitude or the oscillation frequency.
Recent studies \citep{LD10} suggest that the QPO is composed of multiple independent oscillations 
of various duration, but the oscillation frequency is constant over this duration.
A number of methods to couple the QPO with the broadband noise were discussed \citep{Burderi97,LS97,MBS03,IvdK13}.
Similar to the broadband noise, we choose to simulate the QPO from its PSD (narrow Lorentzian) 
using the \citet{TK95} algorithm, with zero mean.

If the QPO is an additive process, the light-curve is a sum of mass accretion rate time-series and the QPO. 
One can imagine this case to be realised if a part of radiating matter varies only due to changing mass accretion rate
and another part (e.g., a blob of matter) varies only due to spinning around a black hole.
However, if the QPOs are produced by the rotation of emission pattern of the entire hot flow \citep{FB07,IDF09,VPI13}, 
the process is multiplicative, as it modulates the accretion rate fluctuations, resulting in  the X-ray light-curve  
\begin{equation}\label{eq:xt_num}
 x(t) = \left[1 + \dot{m}(t)\right]\left[1 + \varepsilon_{\rm x} q(t)\right] -1 ,
\end{equation}
where $q(t)$ accounts for the QPO, $\varepsilon_{\rm x}$ is a positive free parameter.
Optical radiation is a sum of two terms 
\begin{equation}\label{eq:ot_num}
 o(t)=s(t)+r_{\rm ds}d(t), 
\end{equation}
where $r_{\rm ds}$ expresses the relative contribution of the disc and the  synchrotron components.
Power spectra of both $s(t)$ and $d(t)$ light-curves are normalised so that their integrals are equal, thus 
$r_{\rm ds}$ gives the ratio of the power spectra of the disc to that of the synchrotron. 
The synchrotron term is defined as
\begin{equation}\label{eq:st_num}
 s(t) = \left[1 - \dot{m}_{\rm filt}(t)\right]\left[1 + \varepsilon_{\rm o} q(t)\right] -1 ,
\end{equation}
where $\dot{m}_{\rm filt}(t)$ is the accretion rate light-curve, which has high frequencies filtered out,
$\varepsilon_{\rm o}$ is a positive free parameter.
The minus sign in front of $\dot{m}_{\rm filt}(t)$ accounts for the anti-correlation of broadband noise \citep[see details in][]{VPV11}, 
while plus signs in front of $\varepsilon_{\rm x}$ and $\varepsilon_{\rm o}$ aim to have two QPOs in phase.
 Thus both the X-ray and the optical light curves are the superposition of the broadband noise and the QPO variability patterns, 
 mathematically described by equations~(\ref{eq:xt_num}) and (\ref{eq:st_num}).

The reprocessed emission is likely modulated at the QPO frequency as well, however the phase shift and the 
oscillation profiles require detailed knowledge of disc parameters and the observer position \citep[see][]{VP15}.
For the sake of simplicity, we consider only broadband fluctuations of the irradiated disc emission (equation~\ref{eq:disc}).

\begin{figure*}
\centering 
\includegraphics[width=12cm]{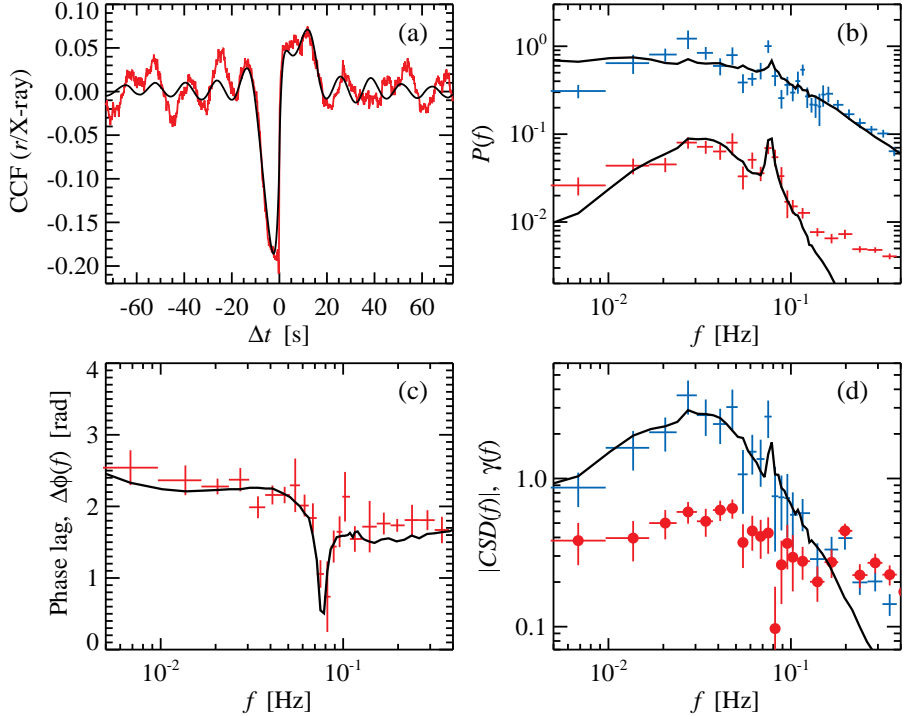} 
\caption{
The observed characteristics of light-curves in $r'$ filter and X-ray (total) band.
The X-ray model characteristics were obtained from the X-ray PSD broadband noise parameters found in \citet[table~5]{DGS09}, coupled 
with the QPO through relation~(\ref{eq:xt_num}).
Theoretical curves were obtained using equations~ (\ref{eq:disc}), (\ref{eq:ot_num}) and (\ref{eq:st_num}).
(a) The observed CCF (red line) and the model (black line).
The model describes well the behaviour of the CCF close to the zero lags and at negative lags.
The divergence at large positive lags is likely caused by additional contribution of the optical QPOs 
coming from irradiation, which we do not account for.
(b) The observed X-ray (blue crosses) and the optical (red crosses) PSDs and 
those given by the model (black solid lines).
QPO is prominent in the optical and tentatively present in the X-ray PSD.
The model well reproduces the shape of optical PSD at frequencies $\lesssim0.2$~Hz.
(c) The observed (red crosses) and the model (black line) phase lags.
(d) The observed coherence $\gamma(f)$ (red bullets with error bars) as well as 
the arbitrary normalised observed (blue crosses) and the model (black line) absolute value of the CSDs. 
}\label{fig:data_model} 
\end{figure*}

\subsection{Results of numerical modelling}

We start our modelling with decomposing the X-ray PSD into Lorentzian functions of the common form 
\begin{equation}
 L_i(f)=\frac{r_i^2 \Delta f_i}{\pi [\Delta f_i^2 + (f-f_i)^2]}, \quad i=0, 1, 2, 
\end{equation}
where $r_i$ describe the normalisations proportional to their contribution to the rms.
The broadband noise is described by two zero-centred Lorentzians, $f_1=f_2=0$ \citep[after][]{NWD99}, with parameters 
found in \citet{DGS09}.
The QPO is described by a Lorentzian with central frequency $f_0=f_{\rm QPO}$ found from the CCF fitting 
and with the width $\Delta f_{\rm QPO}$ found from fitting the phase-lag spectra.
The normalisation of this Lorentzian was chosen such that its peak power is equal to the power of broadband variability at the same frequency, 
$Q(f_{\rm QPO})=\dot M(f_{\rm QPO})$ (or, equivalently, $L_0(f_0)=L_1(f_0)+L_2(f_0)$).
After that, relative importance of the QPO is controlled by coefficients $\varepsilon_{\rm x}$~and~$\varepsilon_{\rm o}$.

\begin{table}
\caption{Parameters of numerical modelling. 
}\label{tab:par}
  \begin{center}
\begin{tabular}{cc}
\hline
Parameter		&	Value		\\
\hline
$r_{\rm ds}$		&	0.85		\\
$t_1$			&	0.1~s		\\
$t_2$			&	5.0~s		\\
$f_{\rm s}$			&	0.05~Hz		\\
$\varepsilon_{\rm x}$	&	0.9		\\
$\varepsilon_{\rm o}$	&	0.8		\\
$f_0$			&	0.078~Hz	\\
$\Delta f_0$		&	0.003~Hz	\\
\hline
\multicolumn{2}{c}{X-ray broadband noise parameters$^a$}  \\
$\Delta f_1$		&	0.095~Hz	\\
$r_1$			&	0.13		\\
$\Delta f_2$		&	1.5~Hz		\\
$r_2$			&	0.09		\\
\hline
$^a$Adopted from \citet{DGS09}.
     \end{tabular}
  \end{center}
\end{table}

Parameters of the model are listed in Table~\ref{tab:par}.
The first four parameters refer to the broadband noise and affect the phase lags, the shape of the optical PSD and the CCF shape 
at small ($|\Delta t|\lesssim10$~s) time lags, while the following four parameters describe the appearance of the QPO:
waves in the CCF at large time lags, narrow spikes in the PSDs, the depth and the width of the dip in the phase lags.

The disc to synchrotron ratio $r_{\rm ds}$ regulates the relative amplitude of the dip to the peak in the CCF.
It simultaneously describes how much the optical PSD is suppressed, relative to the X-ray PSD, at low frequencies. 
The latter is because the disc and the synchrotron emission come (almost) in anti-phase.
Maximal suppression occurs when the synchrotron and disc PSDs are equal \citep[see][]{VPV11}.
The parameter also affects the phase lags: 
the larger is $r_{\rm ds}$ the more phase lags resemble those of the disc (described by an increasing function 
of Fourier frequency), the smaller is $r_{\rm ds}$ the closer phase lags resemble those of the synchrotron term 
(equal to $\pi$ and independent of frequency).

Parameters $t_1$ and $t_2$ define characteristic frequency of suppression of high frequencies in disc PSD and the frequency 
above which the disc phase lags start to substantially differ from zero.
In the CCF, $t_1$ is responsible for the shift of the positive peak and $t_2$ determines its width.

Parameter $f_{\rm s}$ affects the width of the dip in the CCF: the smaller is $f_{\rm s}$ the wider is the dip.
We note that the dip shape generally depends on the shape of the filter function, the latter is related to the physical mechanism
damping the high-frequency fluctuations in the hot flow.
Our choice of $H(f)$, however, was motivated by the fit to the CCF, rather than to its physical nature.

Parameters $\varepsilon_{\rm x}$ and $\varepsilon_{\rm o}$ simultaneously affect the QPO prominence in the PSDs, the wave amplitude  in the CCF and the depth of phase lag drop at $f_{\rm QPO}$. 
The QPO Lorentzian width $\Delta f_{\rm QPO}$ not only determines the 
characteristic timescale on which the oscillations are coherent (i.e. at which time lags oscillations in the CCF are still seen), 
but also influences the width of the dip in the phase-lag spectrum which was used to determine that parameter.

For each set of model parameters we simulate 1500 different realisations of 
the optical and X-ray light curves  
and compute the average model characteristics, which we 
then fit-by-eye to the observed X-ray and optical PSDs, 
the CCF, the  phase lag and the CSD (see Fig.~\ref{fig:data_model}). 
The model parameters that describe the data well are given in Table~\ref{tab:par}.  

\begin{figure*}
\centering 
\includegraphics[width=5.5cm]{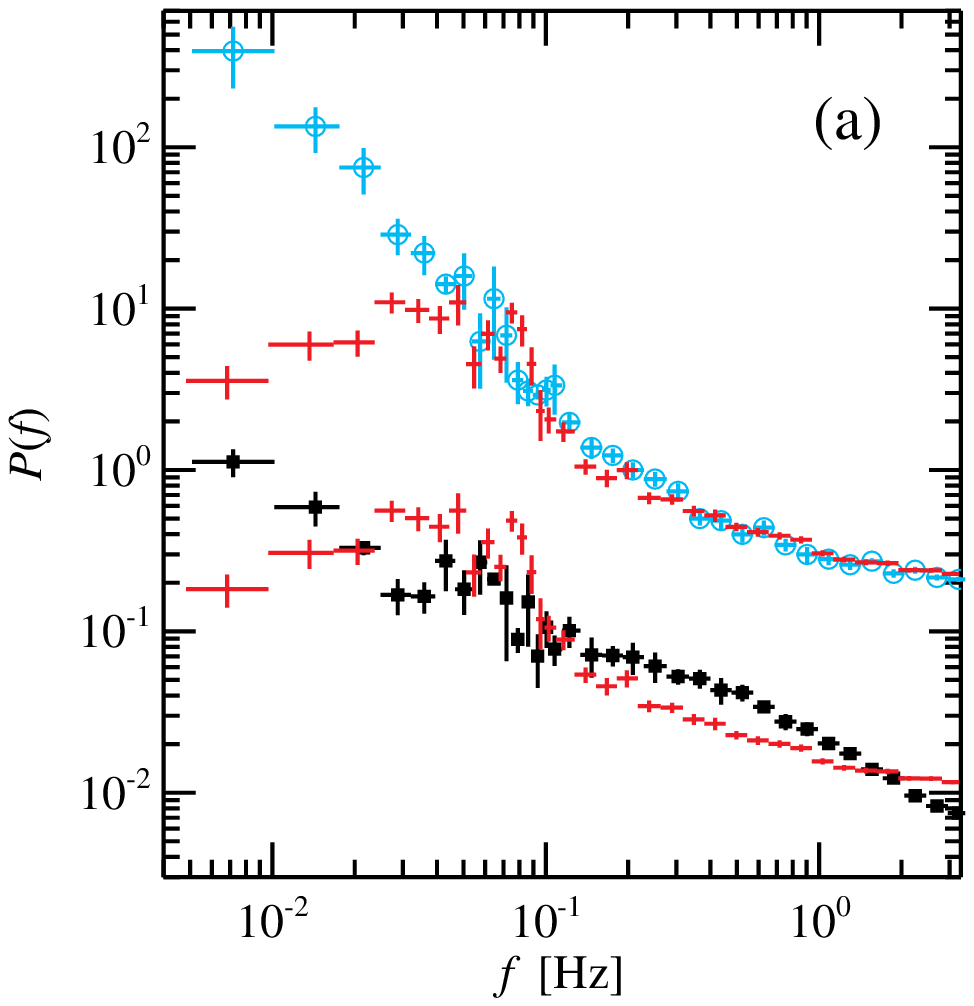} 
\hspace{1cm}
\includegraphics[width=5.5cm]{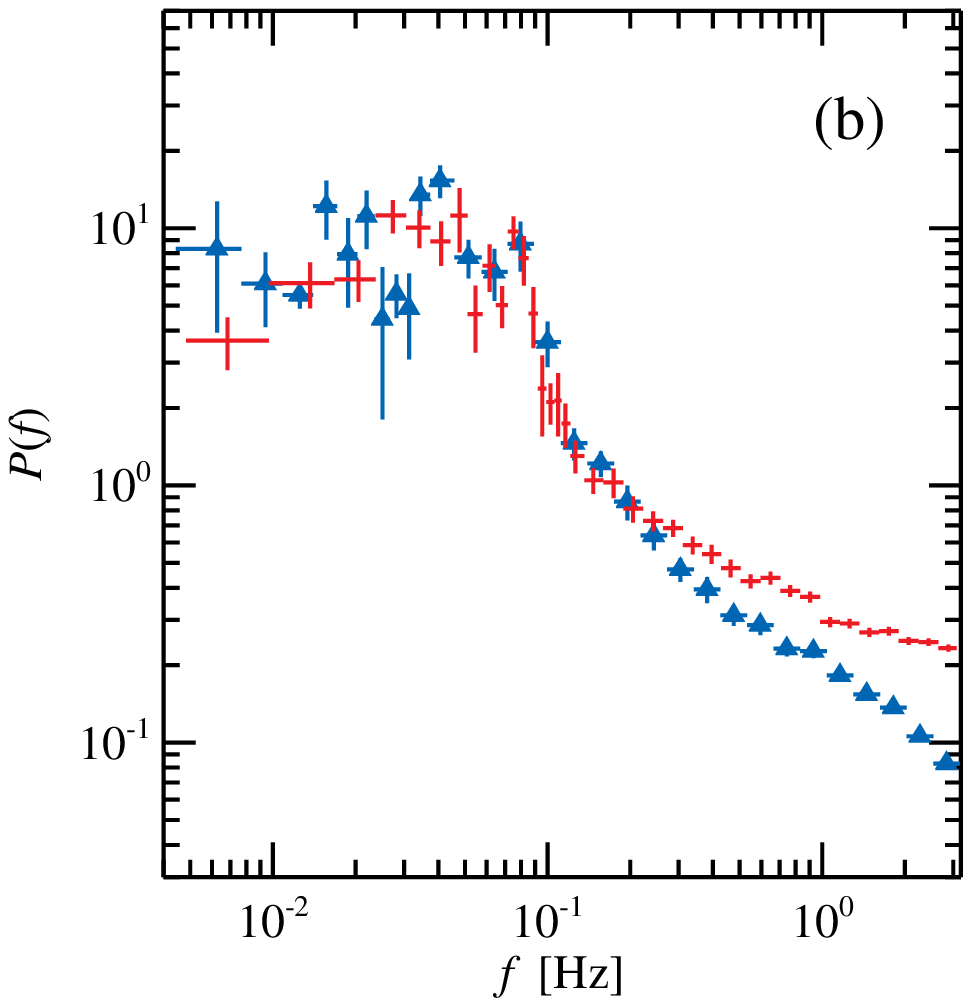} 
\caption{
(a)  The PSDs of SWIFT~J1753.5--0127 (black squares), of the comparison star (cyan circles) and of the ratio of their fluxes 
(red crosses, overplotted on both PSDs) for the night 2007 June 13.
The shape of the ratio-PSD matches the PSD of the comparison star at frequencies $f\gtrsim0.2$~Hz, but is substantially different from the PSD of the source alone.
(b) The PSD of SWIFT~J1753.5--0127 for night 2007 June 18 (blue triangles), as compared to the PSD of the target/comparison star ratio on night 2007 June 13 (red crosses).
The normalisation is chosen such that the two PSDs match at the QPO frequency.
They also match up to frequency $\sim0.2$~Hz and substantially deviate at higher frequencies.
}\label{fig:atm_cont} 
\end{figure*}

The model CCF consists of the correlated aperiodic variability and the correlated QPO, seen as oscillations at large negative 
and positive time lags (see Fig.~\ref{fig:data_model}a).
The anti-correlation at $-10\lesssim\Delta t\lesssim0$~s is described by the synchrotron term, while positive peak of smaller 
amplitude at $0\lesssim\Delta t\lesssim20$~s appears because of the disc term 
(see equations~\ref{eq:response}, \ref{eq:ot_num} and \ref{eq:st_num}).
Interplay of the two optical terms leads to a sharp feature at $\Delta t\cong0$, where their relative role is rapidly changing.
The CCF shape is not well described  at larger positive lags, possibly because of the additional presence of the optical QPO from the disc, 
which we do not account for.
A significant excess at the QPO frequency is seen is also seen in the CSD (Fig.~\ref{fig:data_model}d), 
the Fourier image of the CCF, supporting our finding that the 
oscillations at this frequency in the optical and the X-rays are correlated.  

The X-ray PSD is obviously well described by the model just because we used the parameters from \citet{DGS09} that fit it 
(see Fig.~\ref{fig:data_model}b). 
The optical PSD shape is consistent with  the model at frequencies $f\lesssim0.2$~Hz.
We note that while the X-ray PSD is almost constant at frequencies $f\lesssim0.03$~Hz, the optical power is somewhat suppressed, 
both in the model and in the data.
The suppression again comes from the interplay of two optical components which nearly cancel each other at sufficiently low 
frequencies as they come in anti-phase.

The model phase lags are close to $2$~rad at all frequencies except of the QPO (see Fig.~\ref{fig:data_model}c).
They reflect a joint contribution of the synchrotron ($\Delta\phi=\pi$~rad) and of the disc components 
($\Delta\phi\sim0$ in the frequency range of interest) to the phase lags.
The dip in phase lags at the QPO frequency is also well reproduced.
The phase lags do not reach zero in spite of the fact that the QPOs are in phase, 
because there is still a substantial contribution of the broadband 
variability with $\Delta\phi\sim2$~rad even at the QPO frequency.

The QPO is apparent in the model of optical PSD and there is a hint of the QPO in the model of X-ray PSD, consistent with the data.
The model significantly underpredicts the optical PSD above frequencies $\sim0.2$~Hz.
The cut-off in the model PSD is related to the width of the anti-correlation in the CCF, requiring the synchrotron term to have high 
frequencies damped.
The detailed shape and width of the dip in the CCF depends on the choice of the filter function: the smaller is the characteristic filter 
frequency the wider is the dip and the less power is stored at high frequencies.
At the same time, the disc term does not show much power at high frequencies either.

Some of the model parameters can be  determined from the data with high accuracy. 
For example, the QPO frequency and its width are well defined. 
On the other hand, $t_1$ is rather uncertain. 
We fit the data with $t_1=0.1$~s, while an equally good fit can be achieved with $t_1$ up to 0.5~s. 
The difference can be only seen at high frequencies, which are likely contaminated by the noise
thus are not reliable to prove parameter. 
Although most of the parameters do not have direct physical interpretation, 
parameters $t_1$ and $t_2$ are likely related to the characteristic size of the reprocessing region. 
Interestingly, the value of $t_2=5$~s is consistent with the maximum time delays 
corresponding to the light travel back and forth the accretion disc of radius $6\times10^{10}$~cm 
expected in this system \citep{NVP14}.

\subsection{Nature of the optical variability above 0.1~Hz}\label{sect:discus}

Although the presented model describes rather well  
the X-ray and the optical PSD, the CCF, the phase lags and the CSD, 
there are some deviation in the optical PSD at high frequencies. 
The nature of the high-frequency variability above $>0.1$~Hz is not certain.  
Because the model reproduces the shape of the entire CCF 
(which reflects the correlated signal) and the CSD, the variability is unlikely intrinsic to the source.
The high-frequency noise might be due to the atmosphere, not completely removed by taking the ratio of the BH to the comparison star fluxes.
 
We compare the target and comparison star $r'$-filter PSDs with the PSD of target/comparison star ratio in Fig.~\ref{fig:atm_cont}(a).
We note that the PSD of the ratio is almost identical to the PSD of the comparison star (which reflects the atmospheric noise fluctuations)
at $f\gtrsim0.3$~Hz.
At the same time, the shape of the target PSD is substantially different, demonstrating a hump at these frequencies.
In Fig.~\ref{fig:atm_cont}(b) we compare PSDs of the target and the ratio taken on 2007 June 13 with the target PSD taken on 2007 June 18, 
when the weather conditions were much better.
The PSDs are normalised so that they match at the QPO frequency.
We see that the June 18 target and June 13 ratio PSDs match everywhere below $f\lesssim0.2$~Hz.
Assuming the target PSD did not suffer substantial change within these days, we expect the shape of target June 13 PSD to be different from 
the one demonstrated by the target/comparison star ratio PSD.
We also note drop of the coherence towards high frequencies (see Fig.~\ref{fig:data_model}d), meaning the decrease of power of the 
correlated signal in the PSDs.
These facts support our suggestion that the high-frequency optical power is due to the atmospheric noise rather than being intrinsic to the source.

We note that this noise may also ``veil'' the high-frequency part of the source optical PSD. 
This affects the CCF shape making both the dip and the peak broader, as the high-frequency correlated signal is lost,
and can be the reason for a much narrower dip and peak detected in the 2008 simultaneous data \citep{DSG11}.
From the model point of view, the synchrotron and the disc components may vary on shorter time-scales than those obtained from modelling, 
thus no filter function is needed.
Detailed comparison of the data taken in 2007 and 2008 is needed to clarify this question.

Alternatively, the high-frequency component might be related to synchrotron photons from the hot flow, which are not 
efficiently Comptonized (e.g., from the outer regions of the hot flow which are more transparent to the synchrotron), 
and thus are not correlated with the X-rays.
Given a very low level of radio emission during the outburst history of the source \citep{Soleri10,2015ApJ...808...85T}, 
any jet contribution to the optical wavelengths is doubtful.

\section{Conclusions and outlook}\label{sect:conclus}

We presented the analysis of the simultaneous optical/X-ray data from BH binary SWIFT~J1753.5--0127.
We found presence of the QPO in the X-ray data and demonstrated that it is coupled with the reported 
earlier optical QPO at the same frequency.
We showed that the oscillations are coherent, as the periodic signal is seen in the CCF.
The best-fitting oscillation frequency obtained by approximating the CCF at large lags with the cosine function 
gives $f_{\rm QPO}=0.078$~Hz, the same frequency as the QPO found in the optical PSD.
We found the phase lags are almost independent of the Fourier frequency, $\Delta\phi\sim2$~rad, 
but demonstrated a significant ($4.7\sigma$ and $3.2\sigma$ for $r'$/X-ray and $g'$/X-ray phase lags, respectively) 
decrease at the QPO frequency.
This suggests a separate narrow component is present both in optical and X-ray PSDs, with the phase lags 
significantly lower than those of the broadband noise.
The fact that the CCF was well fitted with the cosine function with zero phase suggests that these 
narrow components are coming in phase ($\Delta \phi=0$).
These findings not only extend the range of X-ray luminosities where the QPOs are present, suggesting those might be
hidden in the noise during the hard state, but also put strong constraints on the origin of optical QPOs,
which has been disputed since their very discovery.

We developed a model which simultaneously explains the shape of the CCF, phase lags and PSDs.
It is based on the temporal behaviour of the hot accretion flow proposed in \citet{VPV11,VPI13}.
In this model, the aperiodic variability in the X-rays is assumed to reflect the propagating fluctuations 
in mass accretion rate. 
The optical emission is composed of two terms: the synchrotron component which varies in anti-phase with the 
X-rays and lacks high-frequency signal (as it is coming from larger radii), and the irradiated disc component 
which is delayed and smeared with respect to the X-rays.
Their interplay results in phase-lags of 2rad, almost independent of frequency.
The X-ray and optical synchrotron broadband variability are coupled to the QPO in a multiplicative way, 
as in the model of the Lense-Thirring precession of the hot flow.
The low value of the phase lags at the QPO frequency is caused by the superposition of the $\sim$2~rad 
lag coming from the aperiodic variability and the QPOs which are coming in phase, $\Delta\phi=0$.

The fact that the QPOs in the optical and X-rays come in phase restricts in the model of \citet[][ see their fig.\,5]{VPI13}
 the orbital parameters of the system, suggesting inclination $i\gtrsim50$\degr.
This is consistent with the constraint $i\gtrsim40$\degr obtained from the prominent 
variability of the optical flux  and of the line equivalent width at the orbital period  \citep{NVP14}.
If indeed $i\gtrsim50$\degr, for the measured mass function $f(M)\lesssim0.95\msun$ \citep{NVP14},
this implies the BH mass of $\lesssim2.5 \msun$, strongly restricting its possible progenitors.

\section*{Acknowledgements}

The work was supported by the Academy of Finland grant 268740 (AV, JP)
and by the Russian Science Foundation grant 14-12-01287 (MR).


\begin{thebibliography}{}
\makeatletter
\relax
\def\mn@urlcharsother{\let\do\@makeother \do\$\do\&\do\#\do\^\do\_\do\%\do\~}
\def\mn@doi{\begingroup\mn@urlcharsother \@ifnextchar [ {\mn@doi@}
  {\mn@doi@[]}}
\def\mn@doi@[#1]#2{\def\@tempa{#1}\ifx\@tempa\@empty \href
  {http://dx.doi.org/#2} {doi:#2}\else \href {http://dx.doi.org/#2} {#1}\fi
  \endgroup}
\def\mn@eprint#1#2{\mn@eprint@#1:#2::\@nil}
\def\mn@eprint@arXiv#1{\href {http://arxiv.org/abs/#1} {{\tt arXiv:#1}}}
\def\mn@eprint@dblp#1{\href {http://dblp.uni-trier.de/rec/bibtex/#1.xml}
  {dblp:#1}}
\def\mn@eprint@#1:#2:#3:#4\@nil{\def\@tempa {#1}\def\@tempb {#2}\def\@tempc
  {#3}\ifx \@tempc \@empty \let \@tempc \@tempb \let \@tempb \@tempa \fi \ifx
  \@tempb \@empty \def\@tempb {arXiv}\fi \@ifundefined
  {mn@eprint@\@tempb}{\@tempb:\@tempc}{\expandafter \expandafter \csname
  mn@eprint@\@tempb\endcsname \expandafter{\@tempc}}}

\bibitem[\protect\citeauthoryear{{Axelsson}, {Borgonovo}  \&
  {Larsson}}{{Axelsson} et~al.}{2005}]{ABL05}
{Axelsson} M.,  {Borgonovo} L.,   {Larsson} S.,  2005, \mn@doi [\aap]
  {10.1051/0004-6361:20042362}, \href
  {http://adsabs.harvard.edu/abs/2005A%26A...438..999A} {438, 999}

\bibitem[\protect\citeauthoryear{{Belloni} \& {Stella}}{{Belloni} \&
  {Stella}}{2014}]{BS14}
{Belloni} T.~M.,  {Stella} L.,  2014, \mn@doi [\ssr]
  {10.1007/s11214-014-0076-0}, \href
  {http://adsabs.harvard.edu/abs/2014SSRv..183...43B} {183, 43}

\bibitem[\protect\citeauthoryear{{Bendat} \& {Piersol}}{{Bendat} \&
  {Piersol}}{1986}]{BP86}
{Bendat} J.~S.,  {Piersol} A.~G.,  1986, Random Data: Analysis and Measurement
  Procedures.
Wiley, New York

\bibitem[\protect\citeauthoryear{{Burderi}, {Robba}, {La Barbera}  \&
  {Guainazzi}}{{Burderi} et~al.}{1997}]{Burderi97}
{Burderi} L.,  {Robba} N.~R.,  {La Barbera} N.,   {Guainazzi} M.,  1997, \apj,
  \href {http://adsabs.harvard.edu/abs/1997ApJ...481..943B} {481, 943}

\bibitem[\protect\citeauthoryear{{Casella}, {Belloni}  \& {Stella}}{{Casella}
  et~al.}{2005}]{CBS05}
{Casella} P.,  {Belloni} T.,   {Stella} L.,  2005, \mn@doi [\apj]
  {10.1086/431174}, \href {http://adsabs.harvard.edu/abs/2005ApJ...629..403C}
  {629, 403}

\bibitem[\protect\citeauthoryear{{Dhillon} et~al.,}{{Dhillon}
  et~al.}{2007}]{Dhillon07}
{Dhillon} V.~S.,  et~al., 2007, \mn@doi [\mnras]
  {10.1111/j.1365-2966.2007.11881.x}, \href
  {http://adsabs.harvard.edu/abs/2007MNRAS.378..825D} {378, 825}

\bibitem[\protect\citeauthoryear{{Done}, {Gierli{\'n}ski}  \& {Kubota}}{{Done}
  et~al.}{2007}]{DGK07}
{Done} C.,  {Gierli{\'n}ski} M.,   {Kubota} A.,  2007, \mn@doi [\aapr]
  {10.1007/s00159-007-0006-1}, \href
  {http://adsabs.harvard.edu/abs/2007A%26ARv..15....1D} {15, 1}

\bibitem[\protect\citeauthoryear{{Durant}, {Gandhi}, {Shahbaz}, {Fabian},
  {Miller}, {Dhillon}  \& {Marsh}}{{Durant} et~al.}{2008}]{DGS08}
{Durant} M.,  {Gandhi} P.,  {Shahbaz} T.,  {Fabian} A.~P.,  {Miller} J.,
  {Dhillon} V.~S.,   {Marsh} T.~R.,  2008, \mn@doi [\apjl] {10.1086/590906},
  \href {http://adsabs.harvard.edu/abs/2008ApJ...682L..45D} {682, L45}

\bibitem[\protect\citeauthoryear{{Durant}, {Gandhi}, {Shahbaz}, {Peralta}  \&
  {Dhillon}}{{Durant} et~al.}{2009}]{DGS09}
{Durant} M.,  {Gandhi} P.,  {Shahbaz} T.,  {Peralta} H.~H.,   {Dhillon} V.~S.,
  2009, \mn@doi [\mnras] {10.1111/j.1365-2966.2008.14044.x}, \href
  {http://adsabs.harvard.edu/abs/2009MNRAS.392..309D} {392, 309}

\bibitem[\protect\citeauthoryear{{Durant} et~al.,}{{Durant}
  et~al.}{2011}]{DSG11}
{Durant} M.,  et~al., 2011, \mn@doi [\mnras]
  {10.1111/j.1365-2966.2010.17604.x}, \href
  {http://adsabs.harvard.edu/abs/2011MNRAS.410.2329D} {410, 2329}

\bibitem[\protect\citeauthoryear{{Fragile}, {Blaes}, {Anninos}  \&
  {Salmonson}}{{Fragile} et~al.}{2007}]{FB07}
{Fragile} P.~C.,  {Blaes} O.~M.,  {Anninos} P.,   {Salmonson} J.~D.,  2007,
  \mn@doi [\apj] {10.1086/521092}, \href
  {http://adsabs.harvard.edu/abs/2007ApJ...668..417F} {668, 417}

\bibitem[\protect\citeauthoryear{{Gandhi} et~al.,}{{Gandhi}
  et~al.}{2008}]{GMD08}
{Gandhi} P.,  et~al., 2008, \mn@doi [\mnras]
  {10.1111/j.1745-3933.2008.00529.x}, \href
  {http://adsabs.harvard.edu/abs/2008MNRAS.390L..29G} {390, L29}

\bibitem[\protect\citeauthoryear{{Gilfanov}}{{Gilfanov}}{2010}]{Gilfanov10}
{Gilfanov} M.,  2010, in {T.~Belloni} ed.,  Lecture Notes in Physics Vol. 794,
  The Jet Paradigm. Springer Verlag, Berlin, p. 17

\bibitem[\protect\citeauthoryear{{Gilfanov}, {Churazov}  \&
  {Revnivtsev}}{{Gilfanov} et~al.}{1999}]{GCR99}
{Gilfanov} M.,  {Churazov} E.,   {Revnivtsev} M.,  1999, \aap, \href
  {http://adsabs.harvard.edu/abs/1999A%26A...352..182G} {352, 182}

\bibitem[\protect\citeauthoryear{{Hynes} et~al.,}{{Hynes} et~al.}{2003}]{HHC03}
{Hynes} R.~I.,  et~al., 2003, \mn@doi [\mnras]
  {10.1046/j.1365-8711.2003.06938.x}, \href
  {http://adsabs.harvard.edu/abs/2003MNRAS.345..292H} {345, 292}

\bibitem[\protect\citeauthoryear{{Hynes}, {O'Brien}, {Mullally}  \&
  {Ashcraft}}{{Hynes} et~al.}{2009}]{HBM09}
{Hynes} R.~I.,  {O'Brien} K.,  {Mullally} F.,   {Ashcraft} T.,  2009, \mn@doi
  [\mnras] {10.1111/j.1365-2966.2009.15260.x}, \href
  {http://adsabs.harvard.edu/abs/2009MNRAS.399..281H} {399, 281}

\bibitem[\protect\citeauthoryear{{Ingram} \& {van der Klis}}{{Ingram} \& {van
  der Klis}}{2013}]{IvdK13}
{Ingram} A.,  {van der Klis} M.,  2013, \mn@doi [\mnras]
  {10.1093/mnras/stt1107}, \href
  {http://adsabs.harvard.edu/abs/2013MNRAS.434.1476I} {434, 1476}

\bibitem[\protect\citeauthoryear{{Ingram}, {Done}  \& {Fragile}}{{Ingram}
  et~al.}{2009}]{IDF09}
{Ingram} A.,  {Done} C.,   {Fragile} P.~C.,  2009, \mn@doi [\mnras]
  {10.1111/j.1745-3933.2009.00693.x}, \href
  {http://adsabs.harvard.edu/abs/2009MNRAS.397L.101I} {397, L101}

\bibitem[\protect\citeauthoryear{{Kanbach}, {Straubmeier}, {Spruit}  \&
  {Belloni}}{{Kanbach} et~al.}{2001}]{Kanbach01}
{Kanbach} G.,  {Straubmeier} C.,  {Spruit} H.~C.,   {Belloni} T.,  2001, \nat,
  \href {http://adsabs.harvard.edu/abs/2001Natur.414..180K} {414, 180}

\bibitem[\protect\citeauthoryear{{Lachowicz} \& {Done}}{{Lachowicz} \&
  {Done}}{2010}]{LD10}
{Lachowicz} P.,  {Done} C.,  2010, \mn@doi [\aap]
  {10.1051/0004-6361/200913144}, \href
  {http://adsabs.harvard.edu/abs/2010A%26A...515A..65L} {515, A65}

\bibitem[\protect\citeauthoryear{{Lazzati} \& {Stella}}{{Lazzati} \&
  {Stella}}{1997}]{LS97}
{Lazzati} D.,  {Stella} L.,  1997, \apj, \href
  {http://adsabs.harvard.edu/abs/1997ApJ...476..267L} {476, 267}

\bibitem[\protect\citeauthoryear{{Lyubarskii}}{{Lyubarskii}}{1997}]{Lyub97}
{Lyubarskii} Y.~E.,  1997, \mnras, \href
  {http://adsabs.harvard.edu/abs/1997MNRAS.292..679L} {292, 679}

\bibitem[\protect\citeauthoryear{{Menna}, {Burderi}, {Stella}, {Robba}  \& {van
  der Klis}}{{Menna} et~al.}{2003}]{MBS03}
{Menna} M.~T.,  {Burderi} L.,  {Stella} L.,  {Robba} N.,   {van der Klis} M.,
  2003, \mn@doi [\apj] {10.1086/374588}, \href
  {http://adsabs.harvard.edu/abs/2003ApJ...589..503M} {589, 503}

\bibitem[\protect\citeauthoryear{{Motch}, {Ricketts}, {Page}, {Ilovaisky}  \&
  {Chevalier}}{{Motch} et~al.}{1983}]{Motch83}
{Motch} C.,  {Ricketts} M.~J.,  {Page} C.~G.,  {Ilovaisky} S.~A.,   {Chevalier}
  C.,  1983, \aap, \href {http://adsabs.harvard.edu/abs/1983A%26A...119..171M}
  {119, 171}

\bibitem[\protect\citeauthoryear{{Neustroev}, {Veledina}, {Poutanen},
  {Zharikov}, {Tsygankov}, {Sjoberg}  \& {Kajava}}{{Neustroev}
  et~al.}{2014}]{NVP14}
{Neustroev} V.~V.,  {Veledina} A.,  {Poutanen} J.,  {Zharikov} S.~V.,
  {Tsygankov} S.~S.,  {Sjoberg} G.,   {Kajava} J.~J.~E.,  2014, \mn@doi
  [\mnras] {10.1093/mnras/stu1924}, \href
  {http://adsabs.harvard.edu/abs/2014MNRAS.445.2424N} {445, 2424}

\bibitem[\protect\citeauthoryear{{Nowak}, {Wilms}  \& {Dove}}{{Nowak}
  et~al.}{1999}]{NWD99}
{Nowak} M.~A.,  {Wilms} J.,   {Dove} J.~B.,  1999, \mn@doi [\apj]
  {10.1086/307189}, \href {http://adsabs.harvard.edu/abs/1999ApJ...517..355N}
  {517, 355}

\bibitem[\protect\citeauthoryear{{Poutanen}}{{Poutanen}}{1998}]{P98}
{Poutanen} J.,  1998, in {Abramowicz} M.~A.,  {Bj\"ornsson} G.,   {Pringle}
  J.~E.,  eds, Theory of Black Hole Accretion Disks. Cambridge University
  Press, Cambridge, p. 100

\bibitem[\protect\citeauthoryear{{Poutanen} \& {Veledina}}{{Poutanen} \&
  {Veledina}}{2014}]{PV14}
{Poutanen} J.,  {Veledina} A.,  2014, \mn@doi [\ssr]
  {10.1007/s11214-013-0033-3}, \href
  {http://adsabs.harvard.edu/abs/2014SSRv..183...61P} {183, 61}

\bibitem[\protect\citeauthoryear{{Remillard} \& {McClintock}}{{Remillard} \&
  {McClintock}}{2006}]{RM06}
{Remillard} R.~A.,  {McClintock} J.~E.,  2006, \mn@doi [\araa]
  {10.1146/annurev.astro.44.051905.092532}, \href
  {http://adsabs.harvard.edu/abs/2006ARA%26A..44...49R} {44, 49}

\bibitem[\protect\citeauthoryear{{Soleri} et~al.,}{{Soleri}
  et~al.}{2010}]{Soleri10}
{Soleri} P.,  et~al., 2010, \mn@doi [\mnras]
  {10.1111/j.1365-2966.2010.16790.x}, \href
  {http://adsabs.harvard.edu/abs/2010MNRAS.406.1471S} {406, 1471}

\bibitem[\protect\citeauthoryear{{Stella} \& {Vietri}}{{Stella} \&
  {Vietri}}{1998}]{SV98}
{Stella} L.,  {Vietri} M.,  1998, \mn@doi [\apjl] {10.1086/311075}, \href
  {http://adsabs.harvard.edu/abs/1998ApJ...492L..59S} {492, L59}

\bibitem[\protect\citeauthoryear{{Timmer} \& {Koenig}}{{Timmer} \&
  {Koenig}}{1995}]{TK95}
{Timmer} J.,  {Koenig} M.,  1995, \aap, \href
  {http://adsabs.harvard.edu/abs/1995A%26A...300..707T} {300, 707}

\bibitem[\protect\citeauthoryear{{Tomsick} et~al.,}{{Tomsick}
  et~al.}{2015}]{2015ApJ...808...85T}
{Tomsick} J.~A.,  et~al., 2015, \mn@doi [\apj] {10.1088/0004-637X/808/1/85},
  \href {http://adsabs.harvard.edu/abs/2015ApJ...808...85T} {808, 85}
  
\bibitem[\protect\citeauthoryear{{Uttley} \& {Casella}}{{Uttley} \&
  {Casella}}{2014}]{UC14}
{Uttley} P.,  {Casella} P.,  2014, \mn@doi [\ssr] {10.1007/s11214-014-0072-4},
  \href {http://adsabs.harvard.edu/abs/2014SSRv..183..453U} {183, 453}

\bibitem[\protect\citeauthoryear{{Veledina} \& {Poutanen}}{{Veledina} \&
  {Poutanen}}{2015}]{VP15}
{Veledina} A.,  {Poutanen} J.,  2015, \mn@doi [\mnras] {10.1093/mnras/stu2737},
  \href {http://adsabs.harvard.edu/abs/2015MNRAS.448..939V} {448, 939}

\bibitem[\protect\citeauthoryear{{Veledina}, {Poutanen}  \& {Vurm}}{{Veledina}
  et~al.}{2011}]{VPV11}
{Veledina} A.,  {Poutanen} J.,   {Vurm} I.,  2011, \mn@doi [\apjl]
  {10.1088/2041-8205/737/1/L17}, \href
  {http://adsabs.harvard.edu/abs/2011ApJ...737L..17V} {737, L17}

\bibitem[\protect\citeauthoryear{{Veledina}, {Poutanen}  \& {Vurm}}{{Veledina}
  et~al.}{2013a}]{VPV13}
{Veledina} A.,  {Poutanen} J.,   {Vurm} I.,  2013a, \mn@doi [\mnras]
  {10.1093/mnras/stt124}, \href
  {http://adsabs.harvard.edu/abs/2013MNRAS.430.3196V} {430, 3196}

\bibitem[\protect\citeauthoryear{{Veledina}, {Poutanen}  \&
  {Ingram}}{{Veledina} et~al.}{2013b}]{VPI13}
{Veledina} A.,  {Poutanen} J.,   {Ingram} A.,  2013b, \mn@doi [\apj]
  {10.1088/0004-637X/778/2/165}, \href
  {http://adsabs.harvard.edu/abs/2013ApJ...778..165V} {778, 165}

\bibitem[\protect\citeauthoryear{{Wijnands} \& {van der Klis}}{{Wijnands} \&
  {van der Klis}}{1999}]{WvdK99}
{Wijnands} R.,  {van der Klis} M.,  1999, \mn@doi [\apj] {10.1086/306993},
  \href {http://adsabs.harvard.edu/abs/1999ApJ...514..939W} {514, 939}

\bibitem[\protect\citeauthoryear{{Zdziarski} \& {Gierli{\'n}ski}}{{Zdziarski}
  \& {Gierli{\'n}ski}}{2004}]{ZG04}
{Zdziarski} A.~A.,  {Gierli{\'n}ski} M.,  2004, \mn@doi [Progr. Theor. Phys.
  Suppl.] {10.1143/PTPS.155.99}, \href
  {http://adsabs.harvard.edu/abs/2004PThPS.155...99Z} {155, 99}

\makeatother
\end{thebibliography}

\bsp	
\label{lastpage}
\end{document}